\documentclass[aip,sd,amsmath,amssymb,reprint]{revtex4-1}

\usepackage{graphicx}
\usepackage{bm}
\usepackage{color}

\begin{document}


\title{Dynamical stability and low-temperature lattice specific heat of one-dimensional fullerene polymers}
\author{Atsushi Shimizu}
\author{Shota Ono}
\email{shota\_o@gifu-u.ac.jp}
\affiliation{Department of Electrical, Electronic and Computer Engineering, Gifu University, Gifu 501-1193, Japan}

\begin{abstract}
We theoretically investigate the dynamical stability of one-dimensional fullerene polymers by computing the phonon dispersion relations within the atomistic approach. We find that only seven models among 54 models proposed previously [J. Phys. Chem. A {\bf 119}, 3048 (2015)] are dynamically stable. We show that the low temperature specific heat of them is proportional to the square root of the temperature in a wider range of temperature compared to the case of single-walled carbon nanotubes.
\end{abstract}


\maketitle

\section{Introduction}
Phonons in nanocarbons such as single-walled carbon nanotubes (CNTs) and graphene have been studied extensively because they play an important role in understanding the thermodynamics \cite{popov,hepplestone,zimmermann,nika} and thermal transport \cite{zimmermann,marconnet} of those systems. Theoretically, the phonon dynamics has been described within the continuum approach and the atomistic approach. While the former is useful to study the low-energy excitations, that is, the acoustic phonon modes, the latter approach is important not only to obtain the optical phonon frequencies but also to understand the dynamical stability of the system.\cite{ito2009,yao2009}

One-dimensional (1D) fullerene polymers (FPs), a new type of nanocarbons, have been created by electron beam irradiation to the C$_{60}$ crystals.\cite{onoe2003} They can be formed through the generalized Stone-Wales (GSW) transformation \cite{SW} between adjacent C$_{60}$ molecules, while the geometry of the coalesced region has not been characterized. A wide variety of interesting phenomena including the photoexcited carrier relaxation,\cite{toda,ono2012,ono2014} the infrared anomaly,\cite{onoe2010} and the anomalous low-temperature resistivity \cite{ryuzaki} have been observed, while a unified interpretation has not been settled. 

Within the first approximation, where the atomic configuration around the coalesced region between the adjacent C$_{60}$ molecules are smeared out, the 1DFPs are regarded as a hollow nanotube with the radius periodically modulated along the tube axis.\cite{onoe2012} One of the authors has investigated the vibrational properties of the 1DFPs within the framework of the continuum elasticity theory.\cite{ono2011} It has been shown that the periodic corrugation induces the zone-folding on the dispersion curves. 

Recently, more than 50 types of the 1DFPs have been proposed through density-functional theory calculations.\cite{noda2015} Depending on the geometry of the coalesced region, the electronic band structure in the 1DFPs shows a gapless property or a finite band gap. However, the phonon band structure, related to the dynamical stability, of the 1DFPs has not been investigated. This is because there are 60 carbon atoms in a unit cell, which makes such a calculation a tedious task to be performed. Since Lindsay {\it et al}. proposed an optimized Tersoff potential for nanocarbons,\cite{lindsay} the lattice dynamics in various systems have been investigated.\cite{marconnet,haskins,koukaras} The use of such an empirical potential would enable us to investigate the phonons of 1DFPs. 

In this paper, using the optimized Tersoff potential, we systematically investigate the dynamical stability of 1DFPs by computing the phonon dispersion relations. We demonstrate that only seven models among 54 models are dynamically stable. It is shown that the specific heat of them is proportional to the square root of the temperature $T$ below 10 K. Such a $T$ range observed in 1DFPs is wider than that observed in single-walled CNTs. 
 
\section{Basic concepts}
We briefly provide some elements of lattice dynamics and formula for the lattice specific heat in solids.\cite{ziman} Let $\bm{R}(s,\bm{l})$ be the $s$th atom position in a unit cell specified by the lattice vector $\bm{l}$. The displacement of the atom from the equilibrium position is $\bm{u}(s,\bm{l})$. The $\alpha$ component is denoted as $u_\alpha(s,\bm{l})$. The motion of equation for $u_\alpha(s,\bm{l})$ is given by
\begin{eqnarray}
 M_s \frac{\partial^2}{\partial t^2} u_\alpha(s,\bm{l}) 
 = - \sum_{s',\bm{l}',\alpha'}
D_{\alpha,\alpha'}^{s,s'}(\bm{l},\bm{l}')
u_{\alpha'}(s',\bm{l}'),
\end{eqnarray}
where $M_s$ is the mass of the $s$th atom. The dynamical matrix is defined by
\begin{eqnarray}
D_{\alpha,\alpha'}^{s,s'}(\bm{l},\bm{l}') =
 \frac{\partial^2 V}{\partial u_\alpha(s,\bm{l}) u_{\alpha'}(s',\bm{l}')} \Big\vert_0,
 \label{eq:dynmat}
\end{eqnarray}
where $V$ is the potential energy. The derivative is evaluated at the equilibrium point. Assuming  
\begin{eqnarray}
u_{\alpha\bm{q}}(s,\bm{l}) = \epsilon_{s\alpha}(\bm{q})
e^{i(\bm{q}\cdot \bm{l} - \omega t)},
\end{eqnarray}
where $\bm{q}=(q_x,q_y,q_z)$ and $\omega$ are the wavevector and the frequency, respectively, one obtains the eigenvalue equation
\begin{eqnarray}
\sum_{s',\alpha'}
\left[
\tilde{D}_{\alpha,\alpha'}^{s,s'}(\bm{q})
- \omega^2 M_s \delta_{s,s'} \delta_{\alpha,\alpha'} 
\right]
\epsilon_{s'\alpha'}(\bm{q}) =0,
\end{eqnarray}
where $\tilde{D}(\bm{q})$ is the Fourier transformation
\begin{eqnarray}
\tilde{D}_{\alpha,\alpha'}^{s,s'}(\bm{q})
&=&
\sum_{\bm{h}}
D_{\alpha,\alpha'}^{s,s'}(\bm{0},\bm{h})
e^{i\bm{q}\cdot \bm{h}}
\end{eqnarray}
with $\bm{h}=\bm{l}'-\bm{l}$. In the present study, the Tersoff potential optimized for nanocarbons \cite{lindsay} is used for $V$ in Eq.~(\ref{eq:dynmat}). The tube axis is assumed to be parallel to $z$-axis, so that the relationship between $\omega$ and $q_z$ yields the phonon dispersion relations $\omega_\gamma (q_z)$, where $\gamma$ is the mode index. 

The lattice specific heat per a unit cell is obtained by using the following expression
\begin{eqnarray}
 C_{\rm lat} = \frac{1}{N} \sum_{\bm{q},\gamma} 
 \hbar\omega_{\gamma}(\bm{q}) \frac{\partial }{\partial T}n_{\rm BE}(\omega_{\gamma}(\bm{q}),T),
 \label{eq:sp}
\end{eqnarray}
where $k_{\rm B}$ is the Boltzmann constant, $N$ is the number of the unit cell, and $n_{\rm BE}(\omega,T)$ is the Bose-Einstein function with $T$.

\section{Results and Discussion}
\subsection{Stable structures}
FP0A model (T3 model in Ref.~\cite{wang}) has a five-fold rotational symmetry around $z$-axis. By performing $n$ times the GSW transformation to the FP0A model, FP$nX$ models are created, where $n=0,1,2,3,4,5$, and 6 and $X=$ A, B, $\cdots$, as described in Ref.~\cite{noda2015}. By referring to the atom position data of the 54 models,\cite{noda2015} we reoptimized their structures by using the optimized Tersoff potential.\cite{lindsay} The structure obtained are almost the same as the original structure, while the lattice constant and the atom positions change slightly.

\begin{figure}[t]
\center
\includegraphics[scale=0.5,clip]{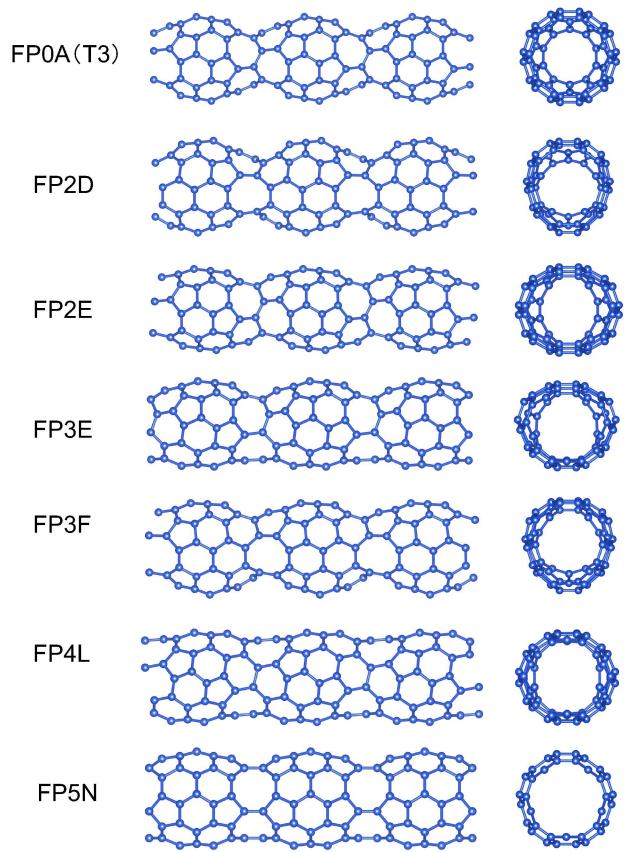}
\caption{\label{fig:structures} Geometry optimized structure of FP0A, FP2D, FP2E, FP3E, FP3F, FP4L, and FP5N models. Side (left) and top (right) views. }
\end{figure}

If the phonon frequency $\omega$ of the FP$nX$ model is positive value for all $q_z$, the FP$nX$ model is dynamically stable. If not, that is, $\omega$ is an imaginary, the model is unstable. We find that among 54 models only seven models are stable: FP0A, FP2D, FP2E, FP3E, FP3F, FP4L, and FP5N, which are shown in Fig.~\ref{fig:structures}. The coalesced region in the FP0A and FP5N models consists of only seven- and eight-membered rings, respectively, endowing those models with a five-fold rotational symmetry around $z$-axis. Since the other models have both seven- and eight-membered rings as well as five- or six-membered rings in the coalesced region, they have a two-fold rotational symmetry or a mirror symmetry.  

\begin{figure}[t]
\center
\includegraphics[scale=0.5,clip]{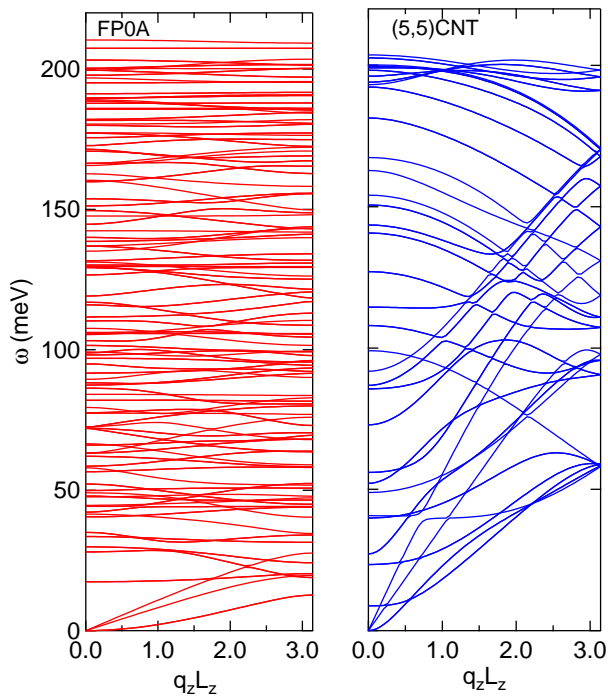}
\caption{\label{fig:band_all} Calculated phonon dispersion curves $\omega_{\gamma}(q_z)$ of FP0A and (5,5)CNT. $L_z$ is 8.80892 \AA \ for FP0A and 2.50615 \AA \ for (5,5)CNT.}
\end{figure}

\subsection{Phonon dispersion relation}
The dispersion curves for FP0A model are shown in Fig.~\ref{fig:band_all}; The case of (5,5)CNT, whose radius is similar to those of FP$nX$ models, is also shown. Since the lattice constant of the former ($L_z=$8.80892 \AA) is more than three times larger than that of the latter ($L_z=$2.50615 \AA), the size of the first Brillouin zone of the former is about one-third that of the latter. Since there are 60 atoms in an unit cell of the FP0A model, 180 dispersion curves are observed. Compared to the case of (5,5)CNT, the dispersion curves in the FP0A model above $\hbar\omega = \hbar\omega_0 = 50$ meV are quite dense, while those below 50 meV are similar. Qualitatively, the periodic corrugation inherent to the FP0A model yields the complex band structure above $\hbar\omega_0$ due to the zone folding effect. The observation of almost flat dispersions above $\omega > \omega_0$ is consistent with the zone-folding scenario. 

\begin{figure}[t]
\center
\includegraphics[scale=0.45,clip]{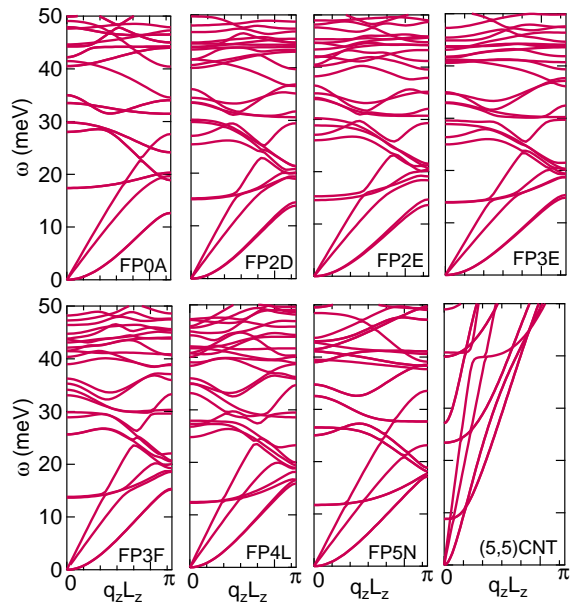}
\caption{\label{fig:band_low} Calculated phonon dispersion curves up to 50 meV for FP0A, FP2D, FP2E, FP3E, FP3F, FP4L, FP5N, and (5,5)CNT.}
\end{figure}

Figure~\ref{fig:band_low} show the phonon band structures of FP0A, FP2D, FP2E, FP3E, FP3F, FP4L, FP5N, and (5,5)CNT up to the phonon energy $\hbar\omega_0$. Due to the tubular structure, all models show four acoustic branches for $q_z\rightarrow 0$: The twisting mode, the longitudinal mode, and the two flexural modes. The two flexural modes are doubly degenerate for FP0A, FP5N, and (5,5)CNT due to the five-rotational symmetry around the tube axis. On the other hand, they are split slightly for the other models because only two-rotational or mirror symmetry are present. As $q_z$ approaches the zone-boundary $\pi /L_z$, the dispersion curves become flat due to the zone-folding effects again. 

The optical mode energy is also influenced by the geometry of the 1DFPs. For example, the energy of the first optical mode at $q_z=0$ decreases with increasing $n$: from 17.4 meV of FP0A to 11.7 meV of FP5N. This is because the averaged tube radius increases with increasing $n$. Moreover, the band repulsion between several dispersion curves is observed due to the mode coupling within $q_zL_z\in (0,\pi)$.

\begin{figure}[ttt]
\center
\includegraphics[scale=0.5,clip]{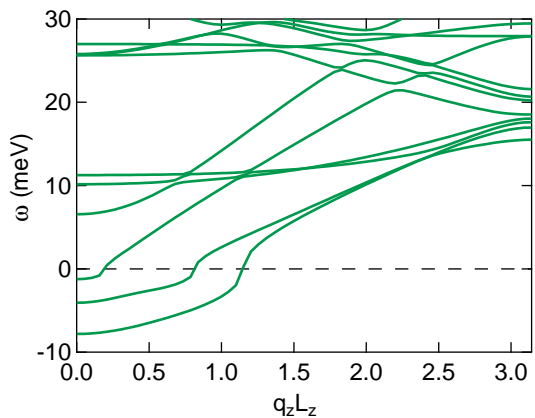}
\caption{\label{fig:band_unstable} Calculated phonon dispersion curves of FP6O model. Imaginary $\omega_{\gamma}(q_z)$s are represented by negative values. }
\end{figure}

Figure \ref{fig:band_unstable} shows the phonon dispersion curves of FP6O model, where imaginary phonon frequencies of unstable modes are represented as negative values. At $q_z=0$, $\hbar\omega=i7.8$, $i4.1$, and $i1.2$ meV. The presence of the imaginary frequencies around $q_z=0$ indicates that the FP6O model is unstable to vibrations with long wavelength. Similar behaviors are observed in the other unstable models. Unfortunately, the relationship between the dynamical stability and the geometry around the coalesced region of the FP$nX$ models is not clear. We postulate that this is related to the symmetry of the system, since most of the unstable models have no rotational and mirror symmetries. 

\subsection{Lattice specific heat}
To better understand the phononic properties of the 1DFPs, we investigate the $T$-dependence of $C_{\rm lat}$ given by Eq.~(\ref{eq:sp}). We assume that the electronic contribution to the specific heat is negligibly small. In general, in the presence of the flexural acoustic modes in 1D systems, where $\omega$ is proportional to the square of $q_{z}$, $C_{\rm lat}$ is proportional to $\sqrt{T}$ as $T$ approaches 0 K. Although such a $T$-dependence of $C_{\rm lat}$ has been predicted in single-walled CNTs\cite{popov,zimmermann} and has been observed experimentally in CNT ropes,\cite{lasjaunias} it holds for millikelvin temperature range only. This is because the other acoustic modes as well as the low-frequency optical modes also start to contribute to $C_{\rm lat}$ as $T$ increases. We can expect that the $\sqrt{T}$ behavior is also observed in 1DFPs even above a few kelvin, since the vibrational energy of the first optical mode of 1DFPs is larger than 10 meV, as shown in Fig.~\ref{fig:band_low}.

\begin{figure}[ttt]
\center
\includegraphics[scale=0.45,clip]{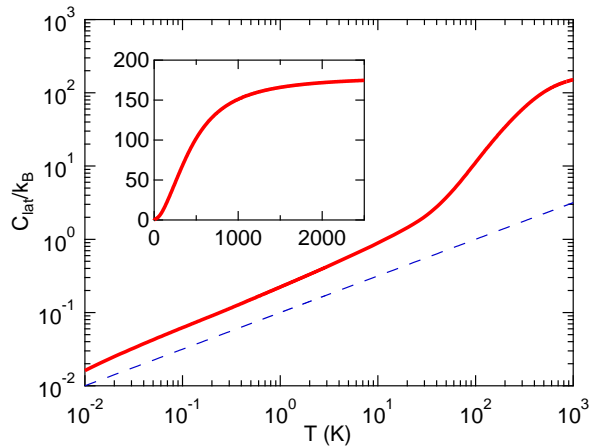}
\caption{\label{fig:specific} Log-log plot of $C_{\rm lat}$, Eq.~(\ref{eq:sp}), of FP0A model as a function of $T$. The dashed line displays the $\sqrt{T}$ behavior of $C_{\rm lat}$. Inset: Normal plot of $C_{\rm lat}$ versus $T$.}
\end{figure}

Figure \ref{fig:specific} shows $C_{\rm lat}$ as a function of $T$ for the case of FP0A. The $C_{\rm lat}$-$T$ curves for FP2D, FP2E, FP3E, FP3F, FP4L, and FP5N models almost overlap with that of FP0A. $C_{\rm lat}$ increases monotonically when $T$ increases. Above $k_{\rm B}T=2000$ K, the value of $C_{\rm lat}$ approaches to $180k_{\rm B}$ (see the inset of Fig.~\ref{fig:specific}), subject to the law of equipartition of energy. As expected, the $\sqrt{T}$ behavior is observed below $T=T_0 \simeq 10$ K. The value of $T_0$ in the FP$nX$ models is quite higher than that in CNTs.\cite{popov,zimmermann,lasjaunias}

\section{Summary}
We have investigated the dynamical stability of 1DFPs by computing the phonon dispersion relations. Among 54 models proposed in Ref.~\cite{noda2015}, seven models, FP0A, FP2D, FP2E, FP3E, FP3F, FP4L, and FP5N, are stable because the phonon frequencies are positive values within the first Brillouin zone. Moreover, we have predicted that the lattice specific heat of those models is proportional to $\sqrt{T}$ below 10 K. We hope that the $\sqrt{T}$ behavior is observed in future experiments. 

\begin{acknowledgments}
The author (SO) would like to thank K. Ohno and Y. Noda for many discussions. This study is supported by a Grant-in-Aid for Young Scientists B (No. 15K17435) from JSPS.
\end{acknowledgments}


\end{document}